# Spatio-Temporal Processing for Automatic Vehicle Detection in Wide-Area Aerial Video

Xin Gao[1] (Member, IEEE), Jeno Szep[1] (Senior Member, IEEE), Pratik Satam[1], Salim Hariri[1] (Senior Member, IEEE), Sundaresh Ram[2] (Member, IEEE), and Jeffrey J. Rodríguez[1] (Senior Member, IEEE)[3]

[1] Department of Electrical and Computer Engineering, The University of Arizona, Tucson, AZ, USA
[2] Department of Radiology, Department of Biomedical Engineering, University of Michigan, Ann Arbor, MI, USA

(This work is partly supported by the Air Force Office of Scientific Research (AFOSR) Dynamic Data-Driven Application Systems (DDDAS) award number FA9550-18-1-0427, National Science Foundation (NSF) research projects NSF-1624668 and NSF-1849113, National Institute of Standards and Technology (NIST) 70NANB18H263 and Department of Energy/National Nuclear Security Administration under Award Number(s) DE-NA0003946.)

**ABSTRACT** Vehicle detection in aerial videos often requires post-processing to eliminate false detections. This paper presents a spatio-temporal processing scheme to improve automatic vehicle detection performance by replacing the thresholding step of existing detection algorithms with multi-neighborhood hysteresis thresholding for foreground pixel classification. The proposed scheme also performs spatial post-processing, which includes morphological opening and closing to shape and prune the detected objects, and temporal post-processing to further reduce false detections. We evaluate the performance of the proposed spatial processing on two local aerial video datasets and one parking vehicle dataset, and the performance of the proposed spatio-temporal processing scheme on five local aerial video datasets and one public dataset. Experimental evaluation shows that the proposed schemes improve vehicle detection performance for each of the nine algorithms when evaluated on seven datasets. Overall, the use of the proposed spatio-temporal processing scheme improves average F-score to above 0.8 and achieves an average reduction of 83.8% in false positives.

## I. INTRODUCTION

Detecting vehicles in low-resolution frames in a wide-area aerial video [1]-[6] can be a challenging task due to imaging issues such as low resolution, camera motion, interlaced videos, and the task-specific detection issues such as low contrast, model diversity, varying degrees of motion and partial occlusion. We have studied a variety of automatic detection and segmentation algorithms in accordance with their related performance evaluation metrics [7]-[18]. Some classical background subtraction algorithms [19]-[21] often assume a video surveillance system using a single, stationary camera. However, the present study aims to track vehicles in an urban traffic scene using a camera aboard a moving helicopter. Therefore, the inconsistent camera motion leads to poor performance for quite a few classical and state-of-the-art approaches to object tracking [20].

When performing multi-stage object detection and classification, it is usually necessary to incorporate a post-processing scheme for reducing false detections. Regarding some of the automatic algorithms [3]-[12], [21]-[30] adapted for vehicle detection using aerial video, the first step of each algorithm applies some form of saliency enhancement, which obtains an initial segmentation of objects versus background. However, the use of these algorithms results in very high error rates, since no post-processing was used to shape the correctly detected objects and drop any incorrect detections. Simply using normalized grayscale thresholds or applying a widely used scheme such as Otsu thresholding [22] for pixel-based classification may result in an unacceptable error rate [13]-[15]. Hence, it is often necessary to augment an automatic detection algorithm with a post-processing scheme.

Research scholars have previously presented several post-processing schemes, i.e., filtered dilation [2], [30], heuristic filtering [3], filtering by shape index [4], sieving and opening [5] as well as sieving and closing [6]. We applied each of these schemes for our vehicle detection task, and recognized their shortcomings as follows [27], [30]. The filtered dilation scheme [2], [30] expands the size of detections, which results in unexpected merging of detected objects. The heuristic filtering [3] reduces false detections only if vehicles have a uniformly distinct aspect ratio and regular size. Filtering by shape index [4] offers limited average F-score improvement; while the strategy of sieving and opening [5] suffers the claim from [11] that objects tend to be wrongly excluded if the background has similar intensity. The sieve and closing scheme [6] uses thresholds for expected vehicle size, which lacks the adaptability to various aerial datasets, while the three-stage post-processing [30] scheme does not exploit spatial connectivity in objects or temporal information for inter-frame correlation.

For the goal of achieving better accuracy when detecting vehicles in aerial videos, we employ spatial post-processing that involves a multi-neighborhood hysteresis thresholding scheme [29] for object segmentation. In the temporal domain, we use two criteria to discard static false detections based on the intersection over union (IoU) and Euclidean distance of the same detection between two adjacent, registered frames. We selected two algorithms that achieved the best overall average F-scores and evaluated performance improvement due to the proposed spatial processing. We determined the best overall size of structuring elements to perform opening and closing using grayscale mathematical morphology to reshape detections and eliminate tiny false objects after thresholding. We also applied the proposed spatio-temporal processing scheme to five algorithms using five local aerial video datasets and verified the performance improvement of post-processing to six algorithms via two public datasets.

The main contributions of our work are presented as below:

i) We derived a spatial processing scheme to reduce false detections when comparing to the ground truth in each frame. This scheme combines 8-neighborhood pixel-weight thresholding and mathematical morphological analysis (opening and closing). Improvement is assessed for nine automatic algorithms using seven aerial video datasets.

ii) We designed a temporal processing scheme using two criteria to further eliminate false detections when using registered frames from an aerial video and then applied the spatio-temporal processing to evaluate its quantitative scores on improving average F-score using consecutive aerial frames in five local datasets and one public dataset.

iii) We used some sensitivity analysis to evaluate the performance of two vehicle detection algorithms using basic information retrieval metrics [13]-[15], [30] under various overlap thresholds for classifying detections, and measured how the size of structuring elements affects the average F-score on four algorithms for spatial processing.

iv) We conducted a variety of experiments to confirm that the proposed spatio-temporal processing significantly drops false positives for all the vehicle detection algorithms under test, using either local or public aerial video datasets.

## II. RELATED WORK

Regarding different types of aerial video datasets with variable grayscale intensity and contrast, we derived a classification policy comprising a thresholding scheme in the spatial domain [27], [29], while its sensitivity may require multiple scaling and shifting factors for specific detection algorithms [29]. Hence, we have improved this thresholding scheme via 8-connected component analysis [6], [27] when thresholding each pixel for object detection, which has increased its robustness and adaptability in practical use. Moreover, we have also improved the step of morphological filtering by using an opening operation for removing tiny false objects and a closing operation for restoring size and shape.

Typically, inter-frame-based algorithms for moving object detection and segmentation in aerial surveillance are mainly categorized as frame differencing, background subtraction, and optical flow approaches [14], [19], [20], each having advantages and disadvantages [20]. For example, $n$-frame cumulative differencing is quite simple and fast with low computational load, but it is sensitive to noise and threshold selection, and it is also vulnerable to failure in detecting large objects having very low contrast. Background subtraction methods can achieve improved object detection accuracy by combining background modeling with frame differencing or employing self-adaptive background models to handle a variable dynamic scene, while the related issues include matching error along with low speed and high memory demands. Optical flow does not require any of the prior information about a traffic scene, however, some challenging issues include multi-optical sources, noise, and shadows, as well as background clutter; meanwhile, its computational complexity makes it difficult to achieve real-time processing, using either global or feature-point optical flow approaches [5]. We implemented some of the schemes mentioned above for aerial vehicle detection, but none of them was a suitable match for post-processing in the temporal domain.

Although machine learning schemes are widely applicable to handle a variety of tasks in the field of digital image analysis, they may have restricted adaptability to resolve our problem on aerial vehicle detection. Since the image resolution and number of available frames is extremely low in several datasets, a training procedure does not improve detection performance. Therefore, our main goal was to investigate how the proposed spatio-temporal processing scheme improves the performance of several detection and segmentation models [7]-[12], [26], [31]-[33] adapted for detecting aerial vehicles.

## III. THE PROPOSED APPROACH: SPATIO-TEMPORAL PROCESSING TO REMOVE FALSE DETECTIONS

The main methodology of our proposed scheme can be divided into two stages. The first stage is spatial processing involving a multi-neighborhood hysteresis thresholding scheme followed by morphological opening and closing operations, applied to each video frame. The second stage is temporal processing, which includes two criteria to drop false detections in each image after the inter-frame comparison. In contrast to our prior work, we removed the binarization step in nine automatic algorithms for detection and segmentation, thereby obtaining adapted algorithms for saliency enhancement. The proposed hysteresis thresholding scheme includes scaling and shifting factors [27], [29] in accordance with the expected distribution of vehicle intensity and contrast in the aerial video datasets; these factors are adjusted (as described in [29]) to achieve the best overall average F-score for each algorithm.

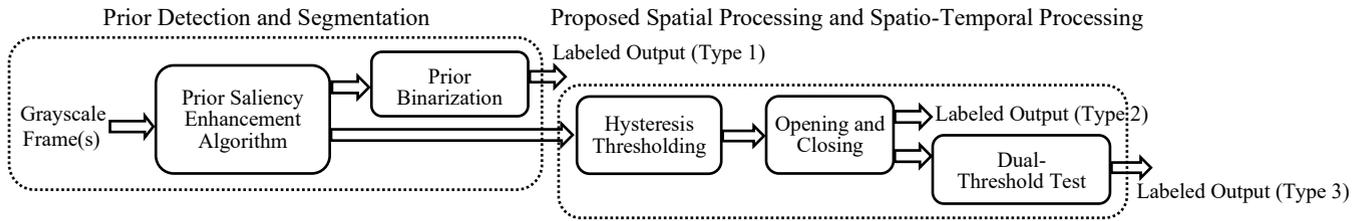

FIG FIGURE 1. Proposed workflow for aerial vehicle detection. Labeled outputs of Type 1, Type 2 or Type 3 represent the detection results of using a prior saliency enhancement algorithm followed by prior binarization, spatial processing or spatio-temporal processing, respectively.

### A. THE GENERAL ARCHITECTURE OF DATA FLOW

The general workflow of our experimental design is depicted in Fig. 1, which comprises an adapted saliency enhancement algorithm followed by the proposed spatial processing and spatio-temporal processing to improve the performance of vehicle detection. Sensitivity analysis was applied to the output of spatial processing, which aims to adjust the overlap threshold for optimal detection of objects and optimize the size of structuring elements for opening and closing to achieve the highest overall average F-score for each algorithm.

### B. SPATIAL PROCESSING

Each grayscale video frame is linearly normalized so the intensity range is in the interval, [0,1], and is then given as input to each of the nine adapted algorithms for saliency enhancement. The enhanced images generated by these algorithms are then fed to our spatial processing scheme, which includes multi-neighborhood hysteresis thresholding followed by morphological filtering.

The multi-neighborhood hysteresis thresholding involves calculating various neighborhood mean values and then subjecting those mean values to hysteresis thresholding. Let $S(r,c)$ represent the segmented image generated by one of the adapted algorithms. Let $L(r,c)$ be the output of the hysteresis thresholding, where each pixel is labeled as foreground or background. We adjust the thresholds (as described in [29]) to achieve the best overall average F-score. For example, multi-neighborhood hysteresis thresholding used in the experiments herein is described as follows:

Step 1: If the pixel value is greater than 5/8, then label the pixel as foreground.

Step 2: If the pixel value is less than 1/8, then label the pixel as background.

Step 3: If the pixel value lies within the interval [1/8, 5/8], then define the neighborhood mean as the mean value of this pixel and its 8 neighbors, and consider the following three cases:

Case 1: If the neighborhood mean is greater than 3/5, then label the pixel as foreground.

Case 2: If the neighborhood mean is less than 1/6, then label the pixel as background.

Case 3: If the neighborhood mean is within the interval [1/6, 3/5], then check the mean value of this pixel and its 4 neighbors, and the mean value of this pixel and its 4 diagonal neighbors. If either of these mean values is greater than 0.5, then label this pixel as foreground; otherwise, label it as background.

Note that the thresholds used above were selected in accordance with the distribution of vehicle intensity and contrast observed within the aerial video datasets used in our experiments. If other datasets have different intensity or contrast – e.g. if vehicles are predominantly dark instead of bright – the thresholds may easily be adjusted accordingly.

Given the fact that some detection errors may persist in complicated traffic scenes even after applying the multi-neighborhood hysteresis thresholding, we perform a follow-up step of using morphological filtering – i.e., an operation of morphological opening to sieve out the trivial false detections, followed up with an operation of morphological closing to connect adjacent small objects, smooth the border of each detection, and fill the tiny holes inside each detection. We use a 2 × 2 square structuring element to perform the required opening operation. For the closing operation, we use a circular disk structuring element whose size varies from 3 × 3 for a low-resolution dataset (e.g., Tucson or Phoenix dataset) to 15 × 15 for a high-resolution dataset (e.g., Tempe dataset), and is selected to achieve the best overall average F-score for each of the saliency enhancement algorithms.

After completing the above spatial processing, connected component analysis is performed, which identifies the set of 8-connected pixels for each detected object.

### C. TEMPORAL PROCESSING

Regarding temporal processing of image sequences, SIFT or SURF descriptors [24] have been used for feature extraction, and a partial intensity invariant feature descriptor (PIIFD) [25] has been adopted for inter-frame registration of raw unregistered data. For single-vehicle detection, a dual-criterion method [23] using motion orientation and object rigidity has been proposed. However, these techniques were not sufficient for our aerial video data, partly due to the wide range (from 20 to several hundred pixels) of vehicle size. For the removal of false detections, we developed a different dual-criterion method, which uses other measures.

Let $A_t(j)$ represent the set of pixels forming detected object $j$ in frame $t$, let $\#(S)$ represent the number of pixels in set $S$, and let $Dist(p,q)$ denote the Euclidean distance between pixels $p$ and $q$ in two consecutive frames. We employ a dual-threshold test [27] to determine if an object is a false detection. Specifically, our temporal processing method is as follows:

Discard detected object $j$ from frame $t$ if there exists a detected object $k$ in frame $t+1$ such that

$$\frac{\#(A_t(j) \cap A_{t+1}(k))}{\#(A_t(j) \cup A_{t+1}(k))} > 75\% \quad (1)$$

and

$$\text{Dist}\left(\text{centroid}(V_t(j)), \text{centroid}(V_t(k))\right) < \delta \quad (2)$$

where we use a distance threshold $\delta = 2$ pixels for the low-resolution datasets and $\delta = 5$ for the high-resolution datasets (i.e., Tempe dataset), respectively. In other words, if an object remains static across consecutive frames, it is discarded.

## IV. EXPERIMENTS AND RESULTS

### A. ALGORITHMS TESTED AND EXPERIMENTAL SETUP

We adapted seven saliency enhancement algorithms by removing the binarization step from previously published detection and segmentation algorithms, which we refer to as the spectral residual (SR) approach [7], frequency tuning (FT) [8], maximum symmetric surround saliency (MSSS) [9], Laplacian pyramid transform (LPT) [10], morphological filtering (MF) [11], variational minimax optimization (VMO) [26], [33] and contextual information saliency (CIS) detection [32]. We also adapted two saliency enhancement algorithms from the multiscale morphological analysis (MMA) [12] and the curvelet-Duda [31] algorithm (Duda).

We implemented two versions on each of the nine algorithms: (1) using our prior binarization methods [27], [45] for simple thresholding, and (2) replacing the binarization step with the proposed spatial-processing or spatio-temporal processing. We refer to these two approaches as "before" and "after" the proposed post-processing in our experimental analysis, respectively. We compare the labeled outputs of the "before" and "after" methods to quantify the performance improvement resulting from the proposed post-processing.

We conducted our experiments using MATLAB R2019b on a Windows PC (Intel Core i7-8500U, 1.80 GHz CPU, 16 GB RAM). The average computation time per frame for each detection algorithm got averaged by the total processing time of each algorithm before and after combining the proposed spatio-temporal processing scheme divided by the total number of frames on the test.

The proposed spatio-temporal processing was evaluated using five local aerial video datasets where vehicles are of various resolutions. We also used two publicly available aerial video datasets for the verification of our approach.

### B. DATASETS

We collected four aerial video datasets (including 9,871 vehicles in total) with very low resolution and one dataset (including 1,114 vehicles) with higher resolution. Several sample frames from each dataset are depicted in Fig. 2. Each frame in The University of Arizona (UA) North Campus dataset has a spatial resolution of $2560 \times 1920$ pixels, while all the other four datasets (Tucson, Phoenix, Tempe Roadway, and UA South Campus) have a spatial resolution of $720 \times 480$ pixels. Traffic lanes were manually cropped from the aerial videos, converted to grayscale, and registered with the other frames. The area of each vehicle varies from 20 pixels to several hundred pixels. Vehicles were manually segmented as a rectangular shape to form the ground-truth segmentation, including thousands of vehicles for each dataset.

We also used two public aerial datasets named as VIVID200 (a subset of 200 consecutive frames with a spatial resolution of $640 \times 480$ pixels, including a total of 1166 vehicles) and a parking dataset (10 aerial images with spatial resolution ranging from $400 \times 800$ to $1600 \times 200$ pixels, including a total of 205 vehicles).

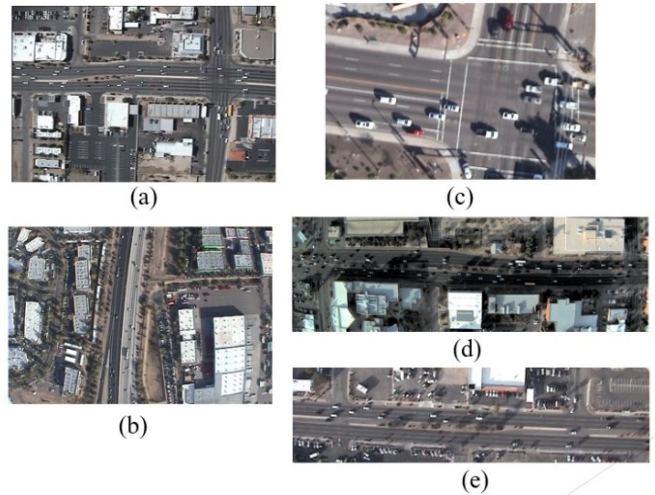

**FIGURE 2.** Sample frames from five aerial video datasets: (a) Tucson, (b) Phoenix, (c) Tempe, (d) UA North, (e) UA South.

### C. PERFORMANCE METRICS

To evaluate the performance of each detection algorithm, we automatically calculate the overlap between detections and ground-truth (GT) objects. Detections can be classified as follows [13], [27], [30]:

- True positive (TP): the correct detection. If multiple detections intersect the same GT object, only the detection having the largest overlap with GT is counted and marked as a TP. If a single detection intersects multiple GT objects, TP is counted only once.
- False negative (FN): detection failure, indicated by a GT object failing to intersect any detected object.
- Splits (S): if multiple detections touch a single GT object, except for the one having largest overlap with GT (marked as TP), all other touches are counted as Splits.
- Merges (M): if a single detection is associated with multiple GT objects, all other objects except the TP in a row are counted as Merges;
- False positive (FP): incorrect detection, indicated by detection that does not intersect any GT object;

Since there are no true negative (TN) samples based on our classification, TN = 0.

Basic information retrieval (IR) metrics [13]-[15], [30] include precision, recall, F-score, and percentage of wrong classifications (PWC). We use average F-score and PWC as metrics to quantify detection performance. PWC [6], [14], [27]-[30] is defined as the ratio of false detections and missed objects to the sum of detections and missed objects, which is formulated as [6], [30]

$$PWC = \frac{FP + FN}{TP + FN + FP + TN} \quad (3)$$

The biased *F*-measure is used as a weighted harmonic mean of precision and recall with a non-negative weight $\beta$, where $\beta^2$ is typically set as 0.3 for saliency detection [15]:

$$\mathbf{F}_\beta = \frac{(1 + \beta^2)\,(\text{Precision})\,(\text{Recall})}{\beta^2\,\text{Precision} + \text{Recall}} \quad (4)$$

when $\beta = 1$, the $\mathbf{F}_\beta$ measure is identical to *F*-score or $F_1$, the harmonic mean of precision and recall, such that Eq. (4) is simplified as [30]

$$F_1 = \frac{2(TP)}{2(TP) + FP + FN} \quad (5)$$

Let an **Ovlp** matrix denote the overlap between each manual GT object and each detected object. We denote the set of pixels comprising each manual GT object and automatically detected objects as $M_i$ and $A_j$, respectively. Note that if an automatic detection consists of more than one region touching a manual GT object, we record all these overlaps for every possible match in the **Ovlp** matrix. Hence, each element of the **Ovlp** matrix is formulated as [29], [30]

$$\text{Ovlp}_{ij} = \frac{\#(M_i \cap A_j)}{\#(M_i \cup A_j)} \quad (6)$$

We select a threshold $\lambda$ to use when thresholding the overlap ratio between each detection and each object. If there exists a non-zero element in a row of **Ovlp**, just label the highest $\text{Ovlp}_{ij}$ as TP if it is above $\lambda$; otherwise, label it as FN.

### D. RESULTS AND ANALYSIS

The quantitative results were evaluated in the following scenarios: (1) six detection algorithms each combined with the proposed spatial processing scheme; (2) selected two algorithms combined with this scheme with various overlap thresholds; (3) quantitative scores of five algorithms before and after spatio-temporal processing; (4) sensitivity analysis of closing operation versus the size of disk filter for spatial processing for four algorithms; (5) the verification of our proposed scheme for six algorithms using two public aerial video datasets; (6) average computation time for nine algorithms before and after spatio-temporal processing. Among the tests as mentioned above, (1) and (2) were performed on Tucson and Phoenix datasets, (3) represented experimental results obtained from the five aerial video datasets depicted in Fig. 2, (4) and (5) involved analysis using VIVID200 and 10 aerial frames of the parking dataset, while (6) combined the results for each of the nine algorithms before and after our spatio-temporal processing performed in their respective test scenarios.

#### 1) Six Automatic Detection Algorithms Applying the Proposed Spatial Processing

The first set of experiments computed the average F-score of each algorithm before and after combining with proposed spatial-processing. Average F-scores of each algorithm before and after using spatial processing are presented in Table 1, where the numerical results show the highest overall average F-score on the tested images for each dataset. For the Tucson (T) dataset, LPT achieved the best average F-score (0.89) after spatial processing, displaying the most significant improvement. For the Phoenix (P) dataset, after spatial processing, MMA and LPT achieved the highest (0.74) and second-highest average F-score (0.71), respectively; all the six algorithms combined with spatial processing display higher improvements on their average F-scores for the Tucson dataset than for the Phoenix dataset, which is likely due to the appearance model diversity and a variable degree of motion [27], [29].

**TABLE 1.** Average F-score for each algorithm before and after combining with the proposed spatial processing

| Alg. | F-Score | Tucson (T) | Phoenix (P) |
|---|---|---|---|
| SR | Before | 0.63 | 0.46 |
| | After | 0.78 | 0.59 |
| FT | Before | 0.56 | 0.39 |
| | After | 0.87 | 0.58 |
| MSSS | Before | 0.51 | 0.41 |
| | After | 0.82 | 0.63 |
| LPT | Before | 0.24 | 0.20 |
| | After | 0.89 | 0.71 |
| MF | Before | 0.33 | 0.15 |
| | After | 0.76 | 0.45 |
| MMA | Before | 0.56 | 0.52 |
| | After | 0.82 | 0.74 |

The classified detections of each algorithm before and after combining spatial processing are tabulated in Table 2, where the overlap threshold for classifying detections was set to zero ($\lambda = 0$), and the last four rows present quantitative results of average $F_\beta$-score. For each algorithm combined with our proposed spatial processing, the counts of S and FP are obviously decreased compared to the original counts, while the removal of FPs results in a loss of TPs for each algorithm; by column comparison, we see that spatial processing also leads to a distinct reduction of Splits and a slight increase of Merges. For the Tucson dataset, LPT yields the highest average $F_\beta$ score (0.93) and all six algorithms achieve an average $F_\beta$ score higher than 0.75 after spatial processing; for the Phoenix dataset, MMA shows the highest average $F_\beta$ score (0.74), which is the only algorithm with an average $F_\beta$ score higher than 0.70 after spatial processing.

The PWC statistics for all six algorithms before and after combining spatial processing are displayed in Table 3, where the sample mean values for the Tucson and Phoenix datasets are given, along with a 95% confidence interval (CI). By column comparison in this table, we see that the PWC scores of all six algorithms are reduced to below 50% for the Tucson dataset, while only LPT and MMA improved their PWC scores to below 50% for the Phoenix dataset. Hence, we conclude that using the proposed spatial processing, LPT and MMA display the best two overall improvements of PWC among these six detection algorithms for the Tucson and Phoenix datasets.

**TABLE 2.** Classified type of detections and $F_\beta$ score (last two rows) for six algorithms before and after applying the proposed spatial processing

| Detection | | Alg. | SR | FT | MSSS | LPT | MF | MMA |
|---|---|---|---|---|---|---|---|---|
| TP | Before | T | 3575 | 3987 | 3985 | 3894 | 3843 | 3582 |
| | | P | 2511 | 3253 | 3472 | 2927 | 2910 | 3018 |
| | After | T | 3322 | 3704 | 3676 | 3552 | 2733 | 3026 |
| | | P | 2235 | 2113 | 2434 | 2768 | 1808 | 2598 |
| S | Before | T | 2793 | 3433 | 4058 | 4038 | 7704 | 980 |
| | | P | 250 | 1376 | 1849 | 1591 | 2396 | 1377 |
| | After | T | 601 | 817 | 415 | 448 | 222 | 8 |
| | | P | 86 | 75 | 140 | 158 | 99 | 32 |
| M | Before | T | 255 | 3 | 15 | 6 | 14 | 162 |
| | | P | 452 | 130 | 338 | 318 | 59 | 455 |
| | After | T | 295 | 183 | 223 | 36 | 80 | 646 |
| | | P | 465 | 48 | 100 | 453 | 286 | 741 |
| FN | Before | T | 182 | 22 | 12 | 112 | 155 | 235 |
| | | P | 1097 | 677 | 250 | 815 | 1091 | 587 |
| | After | T | 435 | 125 | 113 | 424 | 1199 | 340 |
| | | P | 1360 | 1899 | 1526 | 839 | 1966 | 721 |
| FP | Before | T | 4259 | 6471 | 7785 | 22387 | 16835 | 5651 |
| | | P | 3457 | 7550 | 9501 | 22683 | 20907 | 4960 |
| | After | T | 228 | 1405 | 1492 | 180 | 234 | 823 |
| | | P | 1356 | 1324 | 1803 | 1472 | 1655 | 1007 |
| $F_\beta$ | Before | T | 0.52 | 0.44 | 0.40 | 0.22 | 0.15 | 0.46 |
| | | P | 0.47 | 0.35 | 0.32 | 0.15 | 0.23 | 0.44 |
| | After | T | 0.91 | 0.76 | 0.78 | 0.93 | 0.84 | 0.81 |
| | | P | 0.64 | 0.59 | 0.58 | 0.68 | 0.51 | 0.74 |

**TABLE 3.** PWC score for each detection algorithm before and after combining with the proposed spatial processing

| Alg. | Tucson | | Phoenix | |
|---|---|---|---|---|
| | Before | After | Before | After |
| SR | 57.90 ± 0.97 | 16.64 ± 0.96 | 63.87 ± 0.94 | 54.86 ± 0.85 |
| FT | 61.45 ± 0.84 | 29.23 ± 1.21 | 71.54 ± 0.63 | 60.38 ± 1.30 |
| MSSS | 65.70 ± 0.81 | 29.06 ± 1.17 | 73.54 ± 0.56 | 57.78 ± 0.90 |
| LPT | 82.15 ± 1.52 | 14.53 ± 0.83 | 88.70 ± 0.46 | 45.51 ± 1.78 |
| MF | 79.10 ± 1.55 | 34.50 ± 1.89 | 88.02 ± 0.53 | 66.72 ± 1.32 |
| MMA | 61.19 ± 1.15 | 27.76 ± 1.08 | 64.55 ± 0.66 | 37.28 ± 0.96 |

*2) Selected Two Algorithms Combined with Proposed Spatial Processing with Various Overlap Thresholds*

Because they exhibited the best overall improvement in PWC, we selected LPT and MMA as two candidate algorithms for the performance analysis over a range of overlap thresholds $\lambda$ (used for classifying detections), where the average precision, recall, F-score and $F_\beta$ score ($\beta^2 = 0.3$) of LPT and MMA after spatial processing are depicted in a set of subplots in Fig. 3 for the Tucson dataset and Fig. 4 for the Phoenix dataset, respectively. Fig. 3 illustrates that for the Tucson dataset, as the overlap threshold increases, LPT shows better precision but poorer recall in contrast to MMA; however, MMA displays better F-score than LPT when the overlap threshold is within the range of 8% to 40%; regarding $F_\beta$ score, MMA has slightly worse performance than LPT when the overlap threshold is ranging from 0 to 15%, while it is better than LPT when the overlap threshold is higher than 15%. Fig. 4 shows that for the Phoenix dataset, comparing MMA to LPT, the discrepancy of recall increases when the overlap threshold is higher than 5% despite the similarity of their recalls when the overlap threshold is below 5%. Regarding precision, F-score, and $F_\beta$ values, we observe that MMA exhibits better scores in contrast to those of LPT.

The above experimental results for six vehicle detection algorithms before and after spatial processing demonstrate the effectiveness of our proposed scheme based on basic IR evaluation metrics, although this is sometimes at the cost of increasing missed vehicles. As the next set of experiments shows, temporal processing is also necessary for better utilization of motion information among registered frames.

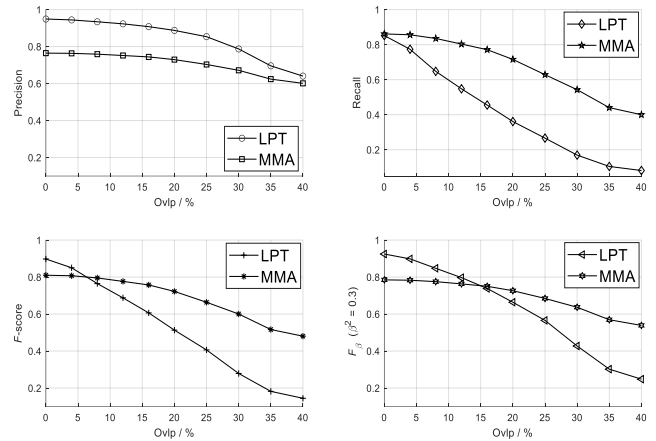

**FIGURE 3.** Performance of LPT and MMA versus overlap threshold (Tucson dataset): precision (top left), recall (top right), F-score (bottom left), and $F_\beta$ value (bottom right).

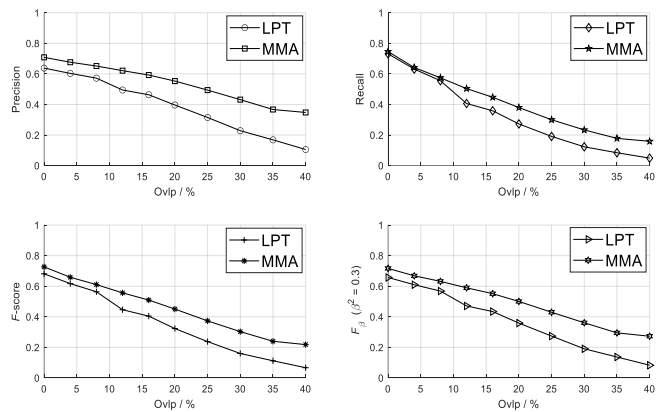

**FIGURE 4.** Performance of LPT and MMA versus overlap threshold (Phoenix dataset): precision (top left), recall (top right), F-score (bottom left), and $F_\beta$ value (bottom right).

*3) Quantitative Evaluation of Five Selected Algorithms Applying the Proposed Spatio-Temporal Processing*

In this subsection, we test how the proposed spatio-temporal processing works for a few selected algorithms using five datasets. After excluding the MF scheme, which displayed the worst overall performance in the experiments discussed above, SR, MSSS, LPT, MMA, and the VMO scheme were selected for further evaluation. We computed the total counts for each type of detections for four low-resolution datasets (Tucson, Phoenix, UA North, and UA South datasets, containing 9871 vehicles in total) using an overlap threshold $\lambda = 0.1$, and for the Tempe dataset (including 1114 vehicles) using an overlap threshold $\lambda = 0.25$ due to higher resolution.

Table 4 provides the classified types of detections for all frames in the four low-resolution datasets (Merged) and the Tempe high-resolution dataset for each algorithm before and after applying the proposed spatio-temporal processing. We observe that the FP count for all five algorithms was reduced by the proposed scheme, with an average reduction of 83.8%. For example, in the case of LPT for the Merged datasets, FP shows a 95.1% reduction; for MSSS with Tempe dataset, FP displays a 96.3% reduction. Regarding the TP count, it varies from a low decrease of 11.3% (LPT) in the Merged four datasets to a high decrease of 21.0% (MSSS) in the Tempe dataset.

**TABLE 4.** Classified type of detections for five algorithms before and after spatio-temporal processing.

| Detection | | Alg. | SR | MSSS | VMO | LPT | MMA |
|---|---|---|---|---|---|---|---|
| TP | Before | Merged | 7439 | 8786 | 8782 | 8323 | 8056 |
| | | Tempe | 1040 | 941 | 1113 | 1103 | 1000 |
| | After | Merged | 6311 | 6940 | 7599 | 7382 | 6698 |
| | | Tempe | 935 | 815 | 952 | 1064 | 867 |
| S | Before | Merged | 3190 | 7162 | 14867 | 6664 | 3781 |
| | | Tempe | 2226 | 6007 | 26015 | 21418 | 2276 |
| | After | Merged | 796 | 982 | 1659 | 805 | 177 |
| | | Tempe | 1087 | 46 | 1978 | 3207 | 340 |
| M | Before | Merged | 786 | 717 | 136 | 324 | 636 |
| | | Tempe | 73 | 173 | 0 | 11 | 114 |
| | After | Merged | 790 | 1110 | 1464 | 511 | 1444 |
| | | Tempe | 0 | 203 | 162 | 12 | 246 |
| FN | Before | Merged | 1646 | 368 | 953 | 1224 | 1179 |
| | | Tempe | 1 | 0 | 1 | 0 | 0 |
| | After | Merged | 2770 | 1821 | 808 | 1978 | 1729 |
| | | Tempe | 179 | 96 | 0 | 38 | 1 |
| FP | Before | Merged | 8515 | 21298 | 58082 | 47097 | 12313 |
| | | Tempe | 8441 | 43397 | 25610 | 69304 | 11087 |
| | After | Merged | 1736 | 8977 | 11641 | 2331 | 2582 |
| | | Tempe | 125 | 1592 | 4743 | 4071 | 4129 |

Average F-score and PWC for each of the five detection algorithms before and after using the proposed scheme are presented in Table 5, along with the 95% confidence interval. After applying spatio-temporal processing, the highest average F-score was observed for SR (0.75), while the lowest was observed for VMO (0.50). Regarding PWC, the highest improvement was observed for LPT, which got reduced by 46.1%; the least improvement was seen in MMA, which was reduced by 27.0%. Among the five algorithms, SR and LPT had PWC below 50% after being combined with the proposed spatio-temporal scheme.

**TABLE 5.** Average F-score and PWC for five algorithms before and after spatio-temporal processing.

| Metric | Alg. | SR | MSSS | VMO | LPT | MMA |
|---|---|---|---|---|---|---|
| F-Score | Before | 0.48 ± 0.01 | 0.23 ± 0.01 | 0.19 ± 0.01 | 0.14 ± 0.01 | 0.42 ± 0.01 |
| | After | 0.75 ± 0.01 | 0.55 ± 0.01 | 0.50 ± 0.01 | 0.67 ± 0.01 | 0.64 ± 0.01 |
| PWC | Before | 68.69 ± 1.53 | 86.99 ± 1.20 | 89.53 ± 1.06 | 92.58 ± 0.79 | 72.25 ± 0.64 |
| | After | 39.90 ± 1.03 | 61.69 ± 1.21 | 66.78 ± 1.29 | 48.92 ± 0.83 | 52.74 ± 1.35 |

### 4) Sensitivity Analysis of Closing Operation vs. Size of Disk Filter for Spatial Processing by Four Algorithms

To investigate adaptability to other datasets, we also tested the proposed spatial processing scheme using two widely recognized public datasets: the VIVID200 dataset and the parking lot dataset.

Given that Li *et al*. [18] investigated the relationship between accuracy and the size of structuring elements for vehicle detection for datasets of different resolutions, we performed a similar analysis using our datasets. We used the average F-score as an evaluation metric to measure four detection algorithms (Duda, CIS, MSSS, and VMO) by using a variable size of structuring elements to perform the morphological closing operations. The radius of the morphological disk filter was varied from 2 to 25. We performed this test using the VIVID200 dataset, and the average F-score for each of the four algorithms is shown in Fig. 5.

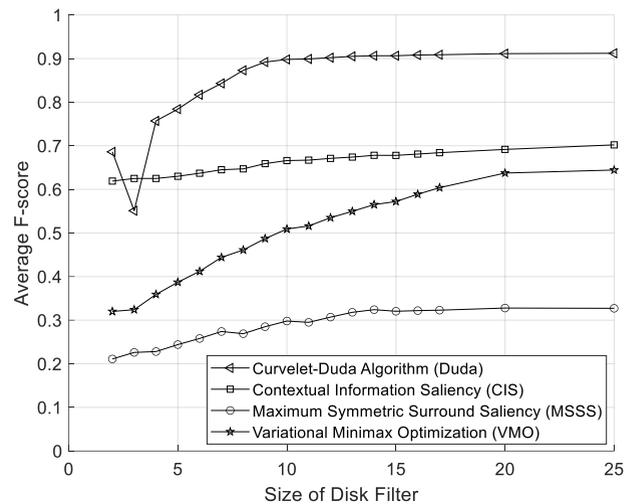

**FIGURE 5.** Average F-score of four algorithms versus the size of the disk filter for morphological closing (VIVID200 dataset).

From Fig. 5, we find that the improved average F-score got close to optimal for Duda, CIS, and MSSS when the filter size reached 10 (where average F-score of VMO equals 0.51); an exception is that when the filter size gets larger, the average F-score of VMO still increases. We applied the proposed spatial processing scheme to each of the four algorithms depicted in Fig. 5 in addition to LPT and MMA, then conducted the performance analysis on ten aerial frames of the parking dataset; due to the image resolution, we used a square 2 × 2 filter for morphological opening followed by a 10 × 10 disk filter for morphological closing. Note that we did not use SR in this set of tests since it had displayed the highest average F-score before temporal processing, and we already had some alternative, adapted algorithms such as MSSS and CIS.

### 5) Performance Verification of Our Approach on Six Detection Algorithms Using Public Aerial Video Datasets

Table 6 presents the classified detections and average F-score for six algorithms before and after applying the spatial processing scheme for the parking lot dataset. From Table 6, we find that all the algorithms except for MSSS and CIS had FPs significantly reduced by over 90%; e.g., VMO displays a 90.5% decrease, and Duda shows a 92.3% reduction. VMO had the best improvement of average F-score and average computation time, while CIS had the lowest values for both measures. Table 7 shows the classified detections for the same six algorithms before and after the proposed spatio-temporal processing for the VIVID200 dataset. In spatial processing, we implemented a square 3 × 3 filter for morphological opening followed by a 10 × 10 disk filter for morphological closing. From Table 7, we observe that except for CIS, all the other five algorithms display more than 90% reduction of FPs; e.g., LPT has a 95.2% reduction and MMA displays a 97.5% reduction. VMO had the highest increase in average F-score, while LPT had the lowest improvement. VMO had the best improvement, with a 95.6% reduction of FPs but a trade-off of 4.63% loss of TPs.

TABLE 6. Performance analysis of six algorithms before and after spatial processing (parking lot dataset).

| Alg. | Time (s) | Classified Detections | | | | | $F_1$ |
|---|---|---|---|---|---|---|---|
| | | TP | S | M | FN | FP | |
| MSSS | Before | 4.50 | 139 | 138 | 3 | 63 | 1114 | 0.26 |
| | After | 1.49 | 114 | 22 | 8 | 83 | 308 | 0.42 |
| VMO | Before | 12.0 | 194 | 1236 | 3 | 8 | 10824 | 0.12 |
| | After | 2.55 | 167 | 96 | 7 | 31 | 1027 | 0.36 |
| CIS | Before | 2.24 | 142 | 60 | 6 | 57 | 374 | 0.39 |
| | After | 1.89 | 102 | 11 | 10 | 93 | 113 | 0.46 |
| Duda | Before | 19.8 | 203 | 1946 | 2 | 0 | 19575 | 0.06 |
| | After | 2.17 | 134 | 119 | 47 | 24 | 1505 | 0.25 |
| LPT | Before | 3.74 | 177 | 491 | 0 | 28 | 3142 | 0.20 |
| | After | 1.65 | 130 | 87 | 16 | 59 | 258 | 0.45 |
| MMA | Before | 3.12 | 170 | 335 | 1 | 34 | 2153 | 0.26 |
| | After | 2.16 | 138 | 76 | 9 | 58 | 202 | 0.52 |

TABLE 7. Performance analysis of four algorithms before and after spatio-temporal processing (VIVID200 dataset).

| Alg. | Time (s) | Classified Detections | | | | | $F_1$ |
|---|---|---|---|---|---|---|---|
| | | TP | S | M | FN | FP | |
| MSSS | Before | 0.66 | 1166 | 2050 | 0 | 0 | 10543 | 0.19 |
| | After | 0.40 | 798 | 197 | 0 | 368 | 591 | 0.63 |
| VMO | Before | 1.06 | 1166 | 3504 | 0 | 0 | 7897 | 0.27 |
| | After | 0.87 | 1112 | 138 | 0 | 54 | 347 | 0.85 |
| CIS | Before | 1.30 | 1112 | 351 | 0 | 54 | 1458 | 0.60 |
| | After | 0.53 | 923 | 69 | 0 | 243 | 351 | 0.76 |
| Duda | Before | 1.65 | 1166 | 3425 | 0 | 0 | 2926 | 0.45 |
| | After | 0.54 | 1166 | 166 | 0 | 0 | 269 | 0.98 |
| LPT | Before | 0.74 | 834 | 3109 | 309 | 23 | 32244 | 0.30 |
| | After | 0.22 | 642 | 12 | 360 | 164 | 1532 | 0.66 |
| MMA | Before | 0.54 | 1166 | 1679 | 0 | 0 | 27160 | 0.08 |
| | After | 0.13 | 625 | 122 | 0 | 541 | 713 | 0.51 |

Visual comparison of six algorithms tabulated in Table 6 and Table 7 before and after applying our proposed scheme is depicted in Fig. 6. Each of the sub-images with size 400 × 400 pixels is selected from the 100th frame of VIVID200 dataset and the 5th frame of the parking lot dataset, where the legends identify each algorithm, the detected vehicles are bounded in color, and the bottom row shows raw data and GT locations. Comparing the results before and after using the proposed scheme, the efficient removal of Splits and FPs is explicitly illustrated for each of the six algorithms.

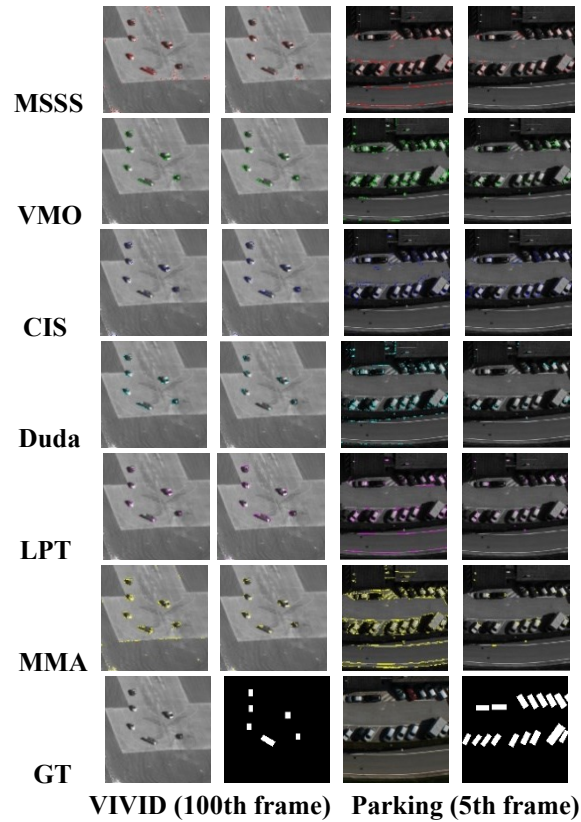

FIGURE 6. Visual comparison of aerial vehicle detection via the six algorithms before and after using the proposed scheme, along with GT (last row).

6) Evaluation of Computation Time for Nine Algorithms Before and After Spatio-Temporal Processing

Table 8 shows the average CPU time for each of the nine detection algorithms before and after using spatio-temporal processing. From column comparison, we observe that the proposed scheme reduces average computation time for each algorithm due to its efficiency in reducing FPs. Duda and MMA display the highest and lowest reduction in CPU time, respectively. CIS shows the fastest run time, while VMO demands the longest computation time.

TABLE 8. Average CPU time of nine algorithms before and after spatio-temporal processing.

| | SR | FT | MSSS | VMO | MF |
|---|---|---|---|---|---|
| Before | 3.10 | 4.94 | 5.76 | 8.36 | 8.17 |
| After | 1.91 | 1.52 | 2.09 | 3.25 | 1.76 |
| | CIS | Duda | LPT | MMA | |
| Before | 1.35 | 2.51 | 5.82 | 3.79 | |
| After | 0.59 | 0.62 | 2.24 | 2.27 | |

V. CONCLUSIONS AND FURTHER WORK

We have derived a spatio-temporal processing scheme on vehicle detection in wide-area aerial imagery, which comprises a multi-neighborhood hysteresis thresholding scheme followed by morphological opening and closing to reduce false detections in the spatial domain, and a temporal processing scheme which further eliminates false detections using two criteria to compare and classify a detection between adjacent frames. The experimental results prove that the proposed scheme is able to achieve performance improvement in vehicle detection accuracy, with the added benefit of reducing average computation time. Performance evaluation of the spatial processing scheme showed that the PWC scores improved the most for the LPT and MMA adapted algorithms, while MMA shows better scores than LPT when the overlap threshold is higher than 15%. Regarding spatio-temporal processing, we observed the efficient removal of FPs for the five algorithms under test. Experiments using public aerial video data (VIVID200 dataset and parking lot dataset) show that the VMO adapted algorithm, combined with our proposed scheme had the best improvement of average F-score.

Some thoughts regarding the future study are summarized as follows: i) Frame differencing can be incorporated as an additional method of removing the false detections, but this was absent in our current study. ii) Although the proposed spatio-temporal processing scheme leads to average F-score improvement for all the involved automatic detection algorithms, the minor loss of TPs and the remaining FPs prevented the achievement of an average F-score higher than 0.9, so this should receive further attention. iii) All seven aerial video datasets contain a limited number of image frames, so larger databases should be established so that adequate training data is available to implement object detection models using some machine learning and deep convolutional neural network (CNN) related schemes [34]-[44]; iv) GPU-based parallel computing has great potential for reducing computation time.

As future work, we plan to consider a parallel design for spatio-temporal processing via a high-performance computer [34] for better utilization of memory and, hence, decreased computation time. We also suggest improving the adapted MMA algorithm by adding a frame differencing method as a pre-processing step, then exploiting better multi-scale feature selection to improve average F-score with reference to some of the latest models [35]-[44] on object detection and their post-processing schemes [45]. Furthermore, we are investigating some of the recent deep learning schemes [46]-[57] for detecting and tracking vehicles [27], [34], [39], [58]-[61] in accordance with the complexity analysis [45], [62]-[69] from the deep CNN-based multi-object detection and segmentation schemes [48]-[51], [53]-[56], [59]-[61], [64]-[66], [70]-[72] applied to wide-area aerial surveillance.

## ACKNOWLEDGMENT
The authors declare no conflict of interest in their study. The authors owe special gratitude to anonymous reviewers for their wonderful suggestions on improving the quality of this paper. The authors are thankful to Prof. Mark Hickman, in the School of Civil Engineering, University of Queensland, Australia, for providing some of the aerial datasets.

## BIOGRAPHIES

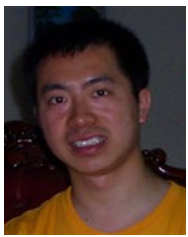

**XIN GAO** (M'10) was born and raised in Taizhou, Jiangsu, China. He received a bachelor (B.E) degree with distinction in information engineering from Nanjing Univ. of Aeronautics and Astronautics (NUAA), China in 2008, a M.S.E.E. degree in electrical engineering from University of Minnesota-Twin Cities, Minneapolis, MN, USA in 2011, a second master's degree in electrical and computer engineering from The University of Arizona, Tucson, AZ, USA, in 2016, and is currently working towards a Ph.D. degree in the same department. In January 2019, he joined the Autonomous Computing Lab. His current research interests mainly include object detection, classification and recognition in aerial imagery as well as visual surveillance in image and video processing.

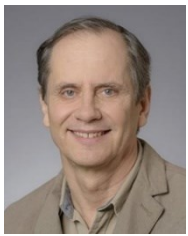

**JENO SZEP** (M'17–SM'20) received his B.S., M.Sc. degrees in Mathematics and Ph.D. degree in Physics from Eötvös Loránd University, Budapest, Hungary in 1979, 1981 and 1985, respectively. He is currently a research scientist in the Department of Electrical and Computer Engineering at The University of Arizona, Tucson, AZ, USA. His research interests mainly include big data analytics, machine learning, statistical signal modelling as well as high-performance computing.

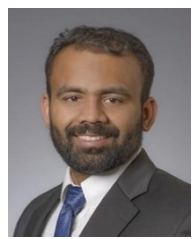

**PRATIK SATAM** received a B.E. degree in electronics and telecommunication engineering from the University of Mumbai, in 2013, a M.S. degree and Ph.D. degree in electrical and computer engineering from The University of Arizona, Tucson, AZ, USA, in 2015 and 2019, respectively. Since 2019, he has been a research assistant professor in the same department at The University of Arizona. His current research interests mainly include autonomic computing, cyber security, cyber resilience, secure critical infrastructures, and cloud security.

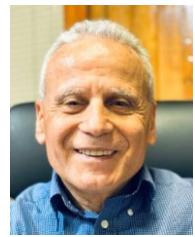

**SALIM HARIRI** (S'81–M'81–SM'05) received the M.Sc. degree from Ohio State University, Columbus, OH, USA, in 1982, and Ph.D. degree in computer engineering from the University of Southern California, in 1986. He is currently a full Professor and the University of Arizona site Director of the NSF-Funded Center for Cloud and Autonomic Computing. He founded the IEEE/ACM Inter-national Symposium on High Performance Distributed Computing, or HPDC. He is also the co-founder of the IEEE/ACM International Conference on Cloud and Autonomic Computing. He has coauthored three books on autonomic computing, parallel and distributed computing, and edited Active Middleware Services, a collection of articles from the second annual AMS Workshop published by Kluwer, in 2000. He serves as the Editor-in-Chief for the scientific journal Cluster Computing, which presents research and applications in parallel processing, distributed computing systems, and computer networks.

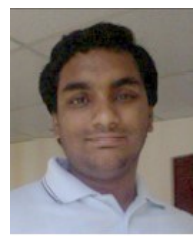

**SUNDARESH RAM** (S'04–M'14) received the B.E. degree in electrical and electronics engineering from the College of Engineering, Anna University, Chennai, India, in 2007, and the M.S. and Ph.D. degrees in electrical and computer engineering from The University of Arizona, Tucson, AZ, USA, in 2010 and 2017, respectively. He was a Post-Doctoral Research Fellow with the School of Electrical and Computer Engineering and the School of Biomedical Engineering, Cornell University, Ithaca, NY, USA, from 2017 to 2018. Dr. Ram is currently a Post-Doctoral Research Fellow with the Department of Radiology, Center for Molecular Imaging, and the Department of Biomedical Engineering, University of Michigan, Ann Arbor, MI, USA. He is also affiliated with the preclinical imaging and computational analysis at Rogel Cancer Center, University of Michigan, Ann Arbor, MI, USA. His research interests include signal processing, image and video processing/analysis, machine learning, computational imaging, and compressive sensing and applications of these methods in the biological sciences, conventional and medical imaging, and data science. He is a member of SIAM and SPIE.

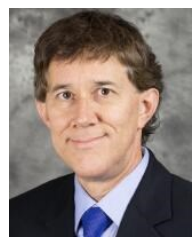

**JEFFREY J. RODRIGUEZ** (M'90–SM'02) received B.S. and Ph.D. degrees from The University of Texas at Austin in 1984 and 1990, respectively, and a master's degree from the Massachusetts Institute of Technology in 1986, all in electrical engineering. In 1990 he joined the faculty at The University of Arizona in Tucson, where he is currently a professor in Electrical & Computer Engineering and Biomedical Engineering, as well as Director of the Signal and Image Laboratory. His research area includes signal/image/video processing and analysis. He served as General Chair of the 2014 and 2016 IEEE Southwest Symp. on Image Analysis and Interpretation, General Chair of the 2007 IEEE Int. Conf. on Image Processing, and on organizing committees for numerous other technical conferences. From 2005 to 2011, he served on the IEEE Signal Processing Society Technical Committee on Image, Video, and Multidimensional Signal Processing. From 2003 to 2008, he was Co-Director of Connection One, a National Science Foundation research center. From 1996 to 2000, he was an Associate Editor of the IEEE Transactions on Image Processing. He is a Senior Member of the IEEE.